\documentclass[useAMS,usegraphicx,usenatbib]{mn2e}

\pagerange{\pageref{firstpage}--\pageref{lastpage}}
\pubyear{2015}

\title
[SSFR in SDSS galaxies: nature or nurture?]
{Nature or nurture? Clues from the distribution of specific star formation rates in SDSS galaxies}

\author[J.~Casado et al.]
{
J.~Casado$^{1,2}$,\thanks{E-mail:javier.casado@uam.es}, Y.~Ascasibar$^{1,2}$, M.~Gavil\'an$^{1,2}$, R.~Terlevich$^{3}$, E.~Terlevich$^{3}$, C.~Hoyos$^{1,2}$, \and A.~I.~D\'{i}az$^{1,2}$\\
$^{1}$Universidad Aut\'onoma de Madrid, 28049 Madrid, Spain\\
$^{2}$Astro-UAM, UAM, Unidad Asociada CSIC\\
$^{3}$Instituto Nacional de Astrof\'{i}sica, \'Optica y Electr\'onica, 72840 Tonantzintla, Puebla, M\'exico
}

\usepackage{color,rotating}

\newcommand{\main}{\emph{main}}
\newcommand{\starforming}{\emph{star-forming}}
\newcommand{\intermediate}{\emph{intermediate}}
\newcommand{\passive}{\emph{passive}}
\newcommand{\agn}{\emph{agn}}
\newcommand{\Ha}{H$\alpha$}
\newcommand{\Hb}{H$\beta$}
\newcommand{\oiii}{[OIII]}
\newcommand{\nii}{[NII]}
\newcommand{\envir}{\ensuremath{R_5/\bar{r}}}
\newcommand{\ratio}{\ensuremath{R_1^2/R_2^2}}
\newcommand{\dd}{{\rm d}}

\newcommand{\figs}{figs}

\usepackage{soul}

\begin{document}

\maketitle

\label{firstpage}

\begin{abstract}
This work investigates the main mechanism(s) that regulate the specific star formation rate (SSFR) in nearby galaxies, cross-correlating two proxies of this quantity -- the equivalent width of the \Ha\ line and the $(u-r)$ colour -- with other physical properties (mass, metallicity, environment, morphology, and the presence of close companions) in a sample of $\sim82500$ galaxies extracted from the Sloan Digital Sky Survey (SDSS).
The existence of a relatively tight `ageing sequence' in the colour-equivalent width plane favours a scenario where the secular conversion of gas into stars (i.e. `\emph{nature}') is the main physical driver of the instantaneous SSFR and the gradual transition from a `chemically primitive' (metal-poor and intensely star-forming) state to a `chemically evolved' (metal-rich and passively evolving) system.
Nevertheless, environmental factors (i.e. `\emph{nurture}') are also important.
In the field, galaxies may be temporarily affected by discrete `quenching' and `rejuvenation' episodes, but such events show little statistical significance in a probabilistic sense, and we find no evidence that galaxy interactions are, on average, a dominant driver of star formation.
Although visually classified mergers tend to display systematically higher EW(\Ha) and bluer $(u-r)$ colours for a given luminosity, most galaxies with high SSFR have uncertain morphologies, which could be due to either internal or external processes.
Field galaxies of early and late morphological types are consistent with the gradual `ageing' scenario, with no obvious signatures of a sudden decrease in their SSFR.
In contrast, star formation is significantly reduced and sometimes completely quenched on a short time scale in dense environments, where many objects are found on a `quenched sequence' in the colour-equivalent width plane.
\end{abstract}


\section{Introduction}
\label{sec_introduction}

Galaxies are known to evolve through a mixture of intrinsic (`nature') processes, such as gas accretion, conversion into stars, and chemical enrichment, as well as through interactions (`nurture') 
with other galaxies.
The observed distribution of their physical properties (e.g. stellar and gas mass, metallicity, SSFR, luminosities and colours) is set by a more or less complex combination of internal and external processes, but their relative importance is still a matter of debate.
In this work, we are interested in the mechanisms that regulate the (specific) star formation rate in nearby galaxies, and, more specifically, the role of dense environments and close interactions on quenching and/or triggering star formation.

Over the last decades, many studies have found that star-forming galaxies (the so-called `blue cloud') are neatly segregated from passively-evolving systems (the `red sequence') in a colour-magnitude diagram \citep[e.g.][]{Tully+82, Strateva+01, Baldry+04, Baldry+06}.
Objects in the intermediate region, known as the `green valley', are typically considered a transition population \citep[e.g.][]{Bell+04, Faber+07, Martin+07, Schiminovich+07, Wyder+07, Mendez+11, Gonsalves+12}.
Many of them display signatures of active galactic nuclei (AGN) in the optical spectrum \citep[e.g.][]{Salim+07, Schawinski+07, Schawinski+10}, and their colours are often interpreted as evidence for a putative `quenching' process that would quickly interrupt any star formation activity.

In this work, we will use the word `quenching' in the sense of a discrete event that instantaneously (or in a very short time scale) truncates the star formation activity in a given galaxy.
In contrast, we will use the term `ageing' to denote a gradual increase in the mass-weighted average age of the stellar population
\begin{equation}
 \tau_*(t) = t - \frac{ \int_0^t t'\, \psi(t')\, \dd t' }{ \int_0^t \psi(t')\, \dd t' }
 \label{eq_age}
\end{equation}
without any sudden change in its star formation rate (SFR) $\psi(t)$.
With this definition, it can be easily shown that galaxies become `older' as they evolve unless their stellar mass increases faster than exponentially (e.g. during a strong starburst).
Therefore, all galaxies are subject to `ageing', whereas they may or may not undergo `quenching' or `rejuvenation burst' episodes.

It has long been known that galaxy colours are strongly correlated with environment, suggesting that `quenching' might be a `nurture' process.
The location of the red sequence is fairly independent on galaxy density \citep{Sandage&Visvanathan78}, and the position of the blue cloud in the colour-magnitude diagram varies only weakly \citep[e.g.][]{Balogh+04_color}.
Most galaxies in the blue cloud display similar SSFRs, varying relatively slowly with mass \citep{Noeske+07, Speagle+14}, and it has been shown \citep[e.g.][]{Balogh+04_Ha, Tanaka+04, Park+07, Wijesinghe+12} that the distribution of \Ha\ equivalent widths for the star-forming population, selected on line intensity and/or colour, is not a strong function of local density.
However, the fraction of red and blue galaxies is a strong function of the environment.

One possible interpretation would be that all galaxies evolve along a `star-forming sequence' until a quenching event would rapidly drive them through the green valley towards the red sequence \citep[see e.g.][]{Peng+10}.
According to these authors, there are two different quenching mechanisms, one related to mass (i.e. `nature') and one related to environment (`nurture').
As galaxies grow in mass, some unidentified process would shut down star formation with a probability that scales linearly with the instantaneous star formation rate (as opposed to e.g. a fixed mass threshold).
`Environment quenching' would be driven by galactic overdensity, and it is found to be roughly independent of stellar mass or cosmic epoch up to $z\sim 1$, consistent with the infall of the galaxy into a larger dark matter halo.
Both mechanisms are independent, and their total effect on the fraction of red galaxies is fully separable.

This kind of scenarios implicitly assume that star-forming and passive galaxies form two distinct groups.
However, some studies question the very existence of such a bimodality in the galaxy population.
Applying data mining techniques, \citet{Ascasibar&Sanchez-Almeida11} found that almost all SDSS galaxies are distributed along a well-defined curve in the  multidimensional space defined by their spectra.
Only optically-bright active galaxies appear as an independent, roughly orthogonal branch that intersects the `main sequence' exactly at the point of the transition between star-forming and passive systems.
The location of such transition is thus well defined, but there is no apparent gap, as far as the optical spectra are concerned.

As recently pointed out by \citet{Schawinski+14}, galaxy morphology may be an important piece of the puzzle.
While early-type galaxies display signatures of rapid quenching, leading to a fast transition through the green valley, late-type galaxies are consistent with slowly-declining star formation rates, with typical time scales of the order of several Gyr.
They do not separate into a blue cloud and a red sequence, but rather span almost the entire $(u-r)$ colour range without any gap or valley in between.
The apparent bimodality in the colour-magnitude diagram arises from the superposition of both morphological populations.

\citet{Schawinski+14} identify early-type galaxies with merger-induced (i.e. nurture) quenching, whereas they advocate for a discrete quenching event in the distant past of late-type galaxies, perhaps associated with reaching a certain halo mass (i.e. nature), that would have shut off the infall of new cold gas.
However, the instantaneous SFR would not drop immediately, but gradually decrease until the cold gas reservoir is exhausted.
Therefore, this process would qualify as `ageing' rather than `quenching' according to our terminology.

Here we would like to statistically quantify the secular evolution of the SSFR as well as the relevance of environment and interactions.
Galaxies in dense environments are subject to a plethora of processes, such as tidal forces, strangulation of the gas supply, ram-pressure stripping, or galaxy harassment \citep[see e.g.][]{Boselli&Gavazzi06} that would suppress the star formation rate, although the exact time scale is still debated \citep[see e.g.][and references therein]{Wijesinghe+12}.

On the other hand, close interactions with neighbouring galaxies can lead to gas compression and trigger star formation \citep[see e.g.][]{Li+08, Ellison+08, Ideue+12}.
Extreme objects such as ultraluminous infrared galaxies are always found to be strongly interacting systems \citep{Sanders+96, Surace+98}.
However, the mechanisms that set the star formation rate at the low-mass end are yet unclear.
Some of these systems show very high SSFRs, often interpreted as massive `rejuvenation' bursts of star formation \citep[e.g.][]{Heckman+98}, but there is little evidence that these episodes are triggered by interactions with nearby neighbours \citep[see e.g.][and references therein]{Fujita+98, Martig+08, Lamastra+13, Lanz+13}.
HII galaxies represent the most extreme objects in this mass range.
They are, in general, blue compact dwarf (BCD) galaxies showing very strong and narrow Balmer emission lines with large equivalent widths \citep{Sargent&Searle70}, and they have the lowest metal content of any star-forming galaxy \citep{Searle&Sargent72, Rosa-Gonzalez+07}.
Observations show that the most luminous HII galaxies tend to live in very low density environments without obvious companions \citep[see e.g.][]{Telles+95, Vilchez95}, although they show irregular morphologies and large velocity dispersions, consistent with a `nurture' origin \citep{Telles+97}.

The main goal of the present work is to investigate the physical properties of star-forming galaxies in the local universe in order to disentangle the contribution of nature and nurture to galaxy ageing, quenching, and rejuvenation.
In particular, we aim to explore the role of stellar mass, local environment, close interactions, and metallicity in setting the SSFR.
Although there is a vast amount of literature on this topic, this work presents one of the most comprehensive studies of all the relevant galaxy properties to date.
It has also sufficient statistics to provide a fair representation of very faint objects, extreme values of the SSFR, and evolutionary phases that last for relatively short times.
We have chosen to use purely observable quantities rather than model-dependent inferences, and we propose new tracers of galactic overdensity and close interactions.
Most importantly, we propose a physical interpretation of our results that is somewhat different from the currently most widely accepted scenarios.

Our selection of a suitable galaxy sample and its physical characterization are described in Section~\ref{sec_observations}. 
The distribution of the different observables we have considered, their correlations, and the implications regarding the mechanism that regulates star formation are discussed 
in Section~\ref{sec_results}. A theoretical interpretation of our results is discussed in Section~\ref{sec_discussion}. Finally, our main conclusions are summarized in Section~\ref{sec_conclusions}.

\section{Observational data}
\label{sec_observations}

\subsection{Sample definition}

In order to proceed with the analysis we require a large sample of galaxies (hereafter \emph{main}) 
that will be used to define neighbours and environment, as well as to select the set of subsamples whose physical properties will be discussed throughout the paper.

All data used in the present analysis are obtained from the SDSS Data Release 7 \citep{DR7} database.
Only objects within the completeness threshold of $m_r<17.77$ and catalogued as galaxies by the spectroscopic pipeline (entries listed in table SpecObj\footnote{http://cas.sdss.org/dr7/en/help/browser/browser.asp}) were chosen to become part of the \textit{main} sample.
Line measurements are obtained from table SpecLine, and absolute magnitudes in the $r$ band, $M_r$, have been computed from the apparent model magnitude\footnote{\tiny http://www.sdss.org/dr7/algorithms/photometry.html$\#$mag\_model} listed in the photometric catalogue (PhotoObjAll table) using the spectroscopic redshift distance.

In addition, we impose the following criteria to select our target galaxies from the \emph{main} sample:
\begin{enumerate}
  \item $ 132.0 < {\rm RA} < 230.0 $ and $ 1.0 < {\rm DEC} < 56.0 $
  \item $m_{\rm r}<17.5$
  \item $ 0.02 < z < 0.07$
\end{enumerate}

\begin{figure}
\centering
\includegraphics[]{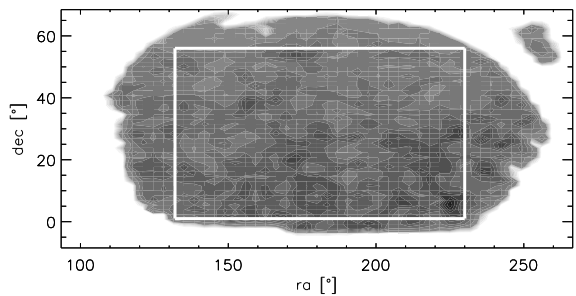}
\caption
{
Spatial distribution of the \main\ sample of galaxies, and region of the sky from which the \emph{star-forming} subsample is selected.
}
\label{fig_footprint}
\end{figure} 

As can be readily seen in Figure~\ref{fig_footprint}, the region of the sky from which the \starforming\ subsample is selected (white square) leaves at least $2^\circ$ ($\sim 3$ and $10.5$~Mpc for $z=0.02$ and $0.07$, respectively) with respect to the main sample footprint, in order to minimize boundary effects when studying interactions and environment.
The restriction in apparent magnitude ($m_{\rm r}<17.5$) ensures that every object with a similar-mass companion would be properly identified.
The upper redshift considered roughly corresponds to an absolute magnitude $M_r<-20.0$, which is approximately the threshold below which most galaxies are actively forming stars, and the lower redshift cut prevents selection effects due to local structure.

\begin{figure}
\centering
\includegraphics[]{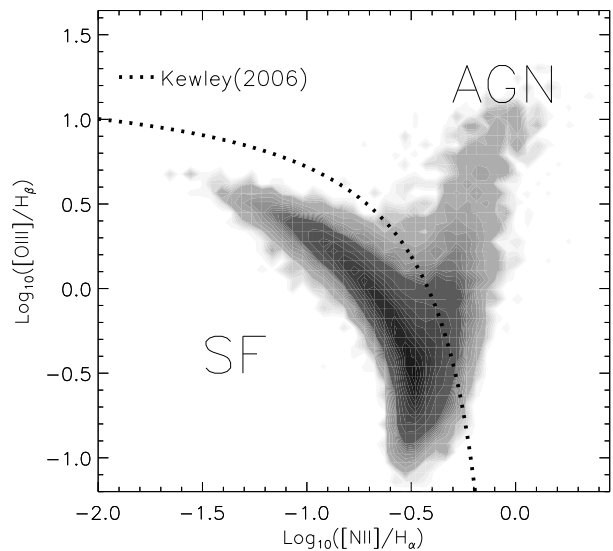}
\caption{Distribution of galaxies with $S/N>2$ measurements of the emission lines in the BPT diagram.
A dotted line shows the \citet{Kewley+06} criterion used to separate star-forming galaxies from AGN.
}
\label{fig_BPT}
\end{figure} 

We then classify our target galaxies into four different subsamples:
\begin{enumerate}
\item Galaxies with signal-to-noise ratio $S/N>2$ in the \Ha, \Hb, \oiii, and \nii\
emission lines, classified as AGN according to the \citet{Kewley+06} criterion based on their position on the BPT diagnostic diagram \citep[see Figure~\ref{fig_BPT}]{BPT} are labelled as \agn\ and will not be considered further.
\item  Galaxies with $S/N>2$ in all the above lines, classified as \starforming\ according to \citet{Kewley+06}, will be referred as such hereafter. For these systems, the imposed $S/N$ ensures that the quality of the spectra is sufficient to estimate the gas-phase metallicity from the O3N2 ratio.
\item Galaxies where \Ha\ is observed in absorption with $S/N>2$ will be referred to as \passive.
\item All other objects are classified as \intermediate.
\end{enumerate}

\subsection{Physical characterization}
\label{sec_characterization}

In order to characterize our galaxy samples in a model-independent fashion we will use direct observables as proxies for their physical properties.

First and foremost, we will use two different indicators of the SSFR, which is the main focus of the present study.
On the one hand, we will consider the equivalent width of the \Ha\ emission line, EW(\Ha), where the line intensity traces the presence of young stars while the underlying continuum is set by the whole star formation history of the galaxy \citep[see e.g.][]{Terlevich+04}.
On the other hand, we will also use the $(u-r)$ colour obtained from the \emph{model} magnitudes, sensitive to the presence of massive, short-lived stars that dominate the bluer spectral bands.
The main differences between both observables are that EW(\Ha) is subject to aperture effects (see below) and that it traces a younger stellar population than $(u-r)$.
Therefore, these two indicators probe the SSFR on different spatial and temporal scales.


In order to describe the intrinsic properties (i.e. `nature') of our objects, throughout the present work we will use the absolute magnitude in the $r$ band, $M_r$, as a proxy for stellar mass, and we will use both terms interchangeably.
To account for their evolutionary state, from the point of view of chemical enrichment, we will use the observational indicator
\begin{equation}
 \rm O3N2 = \log{ \frac{\oiii}{H\beta} } -  \log{ \frac{\nii}{H\alpha} }
\end{equation}
that is well known to decrease monotonically with the gas-phase metallicity \citep{Alloin+79} and thus with stellar-to-gas fraction \citep{Searle&Sargent72, Edmunds90, Ascasibar+14}.

Concerning the effects of `nurture' on the SSFR, we will quantify the environment in terms of the local galaxy overdensity, estimated from the projected distance to the fifth nearest neighbour in the \main\ sample.
We propose the relative distance between the first and second neighbours as a simple indicator for the presence of close companions, and we will interpret galaxy morphology as a possible tracer of galaxy-galaxy interactions.

The definition of distance is a key aspect in the characterization of close interactions and environment.
Spectroscopic redshifts are available for every object in our sample which permits distance calculations in 3D space.
However, redshift measurements are strongly affected by the peculiar velocities of galaxies (especially important in denser environments such as galaxy clusters).
As our galaxy sample covers all kind of environments, this effect will act irregularly and introduce a bias. 
In order to avoid such problems, we base our clustering calculations on projected distance measurements, as it is widely done in the literature \citep[see e.g.][and references therein]{Kaiser87, Cooper+05, Muldrew+12}.

For every target galaxy with redshift $z_{\rm gal}$, we select potential neighbours from the \main\ sample within a range $z_{\rm gal} \pm 0.005$ 
(which corresponds to $\sim 15$~Mpc or $\sim 1500$ km/s velocity difference).
Then, the distance to each object in this redshift shell is calculated as the great-circle distance, i.e. projected onto the surface of a sphere with radius $d_{A}(z)$, where
\begin{equation}
 d_A(z) = \frac{d_L(z)}{(1+z)^2}
\end{equation}
is the angular diameter distance, and $d_L(z)$ corresponds to the luminosity distance at redshift $z$.

Since our \main\ sample is magnitude-limited, the number density of galaxies decreases with redshift as the faintest objects drop out from the sample. 
Hence, the average distances to the neighbours increase with $z$.
We circumvent this problem by normalizing such distances to a characteristic radius
\begin{equation}
\overline{r}(z) = \frac{d_A(z)}{\sqrt{N_{\rm gal}(z\pm0.005)}}
\end{equation}
that is proportional to the average distance between galaxies, assuming they were uniformly distributed over the solid angle $\Delta \Omega$ within the redshift shell $z\pm0.005$ 
(i.e. $\pi\,\overline{r}^2 \propto \Delta\Omega\, d_A^2 / N_{\rm gal}$).
$N_{\rm gal}$ refers to the number of galaxies within the redshift shell under consideration.
Then, we will use the normalized distance to the fifth neighbour, \envir, as a redshift-independent estimator of the local galaxy density 
(i.e. an indicator of the environment where the galaxy lives in).
Galaxies in voids display values of the order of a few, whereas cluster galaxies feature values well below unity.

In order to test for the presence of close companions, we chose to use the ratio between the squared projected distances to the first and second neighbours.
For a uniform random distribution in two dimensions, \ratio\ should be uniformly distributed between 0 and~1, irrespective of the local galaxy density.
A low value of $R_1$ is not in itself indicative of interactions, and it is trivially obtained e.g. for all cluster galaxies.
In contrast, very low values of \ratio\ may mean (in a probabilistic sense) that the first neighbour is much closer than expected for a purely random distribution, 
and thus they trace the fraction of close galaxy pairs in a statistical sample.

Finally, the morphology of a galaxy is also a powerful indicator of its dynamical and merger history.
Thanks to the Galaxy Zoo 1 citizen science project \citep{Lintott+08}, almost every galaxy in the SDSS spectroscopic catalogue has a visual morphology classification.
From the votes of the `citizen scientists', debiased from redshift-dependent resolution effects that may blur galactic features (such as spiral arms) in SDSS images, galaxies are classified as \emph{ellipticals} or \emph{spirals} (clockwise, anti-clockwise and edge on) if either category receives at least 80\% of the votes.
Objects that do not reach the 80\% threshold are designed as \emph{uncertain}, including a fraction of merging systems.
According to \citet{Darg+10a}, a visual merger identification can be considered reliable when it receives more than 40\% of the votes.
Based on these criteria, we label every object in our sample as \emph{early-type}, \emph{late-type}, \emph{uncertain} (excluding mergers) or \emph{merger}.

\begin{figure}
\centering
\includegraphics[width=.45\textwidth]{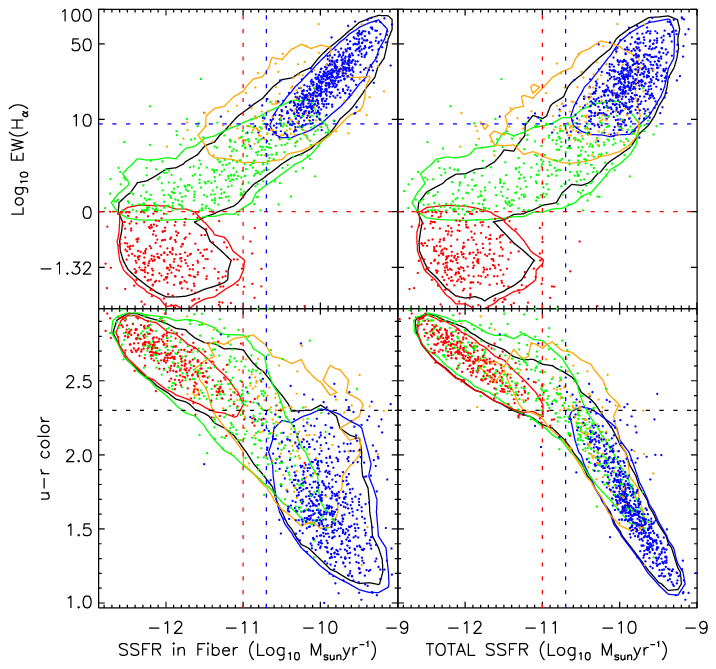}\\[-5mm]
\caption
{
Comparison between model SSFR \citep{Brinchmann+04} and our observational proxies.
Left panels correspond to the SSFR within the fibre, and right panels show the total SSFR.
Top and bottom panels show \Ha\ equivalent width and $(u-r)$ colour, respectively.
Contours in blue, red, green, and orange correspond to the area that encloses $90$\% of the \starforming, \passive, \intermediate, and \agn\ subsamples.
Points correspond to a random selection of 1500 target galaxies, and they are similarly coloured according to the subsample they belong to.
}
\label{fig_comparison_SSFR}
\end{figure} 

\subsection{Observational biases}
\label{sec_bias}

Although our physical characterization is intended to be as objective as possible, it is not completely free from observational biases, some of them inherent to the selection function of the spectroscopic SDSS.

For instance, the minimum fibre separation in the detector implies that galaxies in the spectroscopic catalogue can only be as close $\sim 55$~arcsec in the sky (physical distances of $\sim 22.5$ and $\sim 74$~kpc at our redshift limits of $z=0.02$ and $0.07$, respectively).
This might introduce some bias in the distance to the fifth neighbour, and it poses a serious problem for the identification of close pairs \citep[see e.g.][and references therein]{Behroozi+15}.

In addition, most of the quantities we consider are based on astrometric and/or photometric measurements, but aperture effects may be an issue for spectroscopic observables such as the equivalent width of the \Ha\ line or the O3N2 ratio.
The aperture of the SDSS fibre (3~arcsec in diameter) corresponds to $\sim1-4$~kpc at $z=0.02-0.07$, and therefore these measurements may not always be representative of the whole galaxy.

Aperture bias has been thoroughly discussed in the literature \citep[e.g.][]{Brinchmann+04, Iglesias-Paramo+13}, and models that take into account the photometry outside the fibre have been proposed in order to correct the spectroscopic SSFR from aperture effects.
In Figure~\ref{fig_comparison_SSFR} we compare the direct measurements of EW(\Ha) and $(u-r)$ with the aperture-corrected SSFR inferred from the \citet{Brinchmann+04} prescription, as implemented in the MPA-JHU pipeline\footnote{http://www.mpa-garching.mpg.de/SDSS/DR7/}.

Equivalent width is tightly correlated with the model SSFR in the area covered by the fibre (top left panel), at least as far as \starforming\ and \intermediate\ galaxies are concerned.
For \passive\ galaxies, the \Ha\ absorption line is actually more prominent in systems with higher SSFR due to a larger contribution of A stars to their optical spectrum.
The aperture correction proposed by \citet{Brinchmann+04} does not change these general trends, but it broadens the correlation with the extrapolated SSFR, as shown on the top right panel.
$(u-r)$ colour, on the other hand, is an excellent proxy for the model SSFR over the whole galaxy (bottom right panel), and it is well correlated with the theoretical estimate inside the fibre, even for \passive\ galaxies (bottom left panel).

The distribution of \starforming, \intermediate, and \passive\ galaxies in Figure~\ref{fig_comparison_SSFR} clearly shows that our sample selection criteria are roughly equivalent to fixed thresholds in SSFR, colour, or equivalent width.
The latter arises from the minimum signal-to-noise imposed, which is dominated by the contrast (i.e. the equivalent width rather than the absolute luminosity) of the \Hb\ line with respect to the underlying continuum.
Our $S/N>2$ criterion, and thus our definition of the \starforming\ subsample, is in practical terms very similar to EW(\Ha)$>$10~\AA, $(u-r)<2.3$, or SSFR$>3\times 10^{-11}$~yr$^{-1}$, which may not be statistically representative of the overall galaxy population and introduce a significant source of bias.

\emph{Passive} galaxies typically populate the range $(u-r)>2.3$ and SSFR~$<10^{-11}$~yr$^{-1}$.
The \intermediate\ subsample is presumably dominated by emission-line objects with EW(\Ha)~$<10$~\AA, most likely weakly star forming galaxies, but, given the low signal-to-noise ratio, it also includes an uncertain fraction of absorption-line systems and AGN.
We have made no attempt to correct for such effects.

Finally, we have also assumed that the value of O3N2 within the spectroscopic fibre is representative of the whole galaxy.
This measurement does not account for radial abundance gradients, and (assuming the fibre is placed at the centre of the galaxy) it would be biased high for large spirals.
The precise impact of aperture bias on our results concerning gas-phase metallicity is hard to predict, but the observed values of the radial abundance gradients in CALIFA disk galaxies suggest a variation of the order of $\sim -0.2$~dex between 0.3 and 2 effective radii \citep{Sanchez+14}.

\begin{figure*}
\centering
\includegraphics[width=\textwidth]{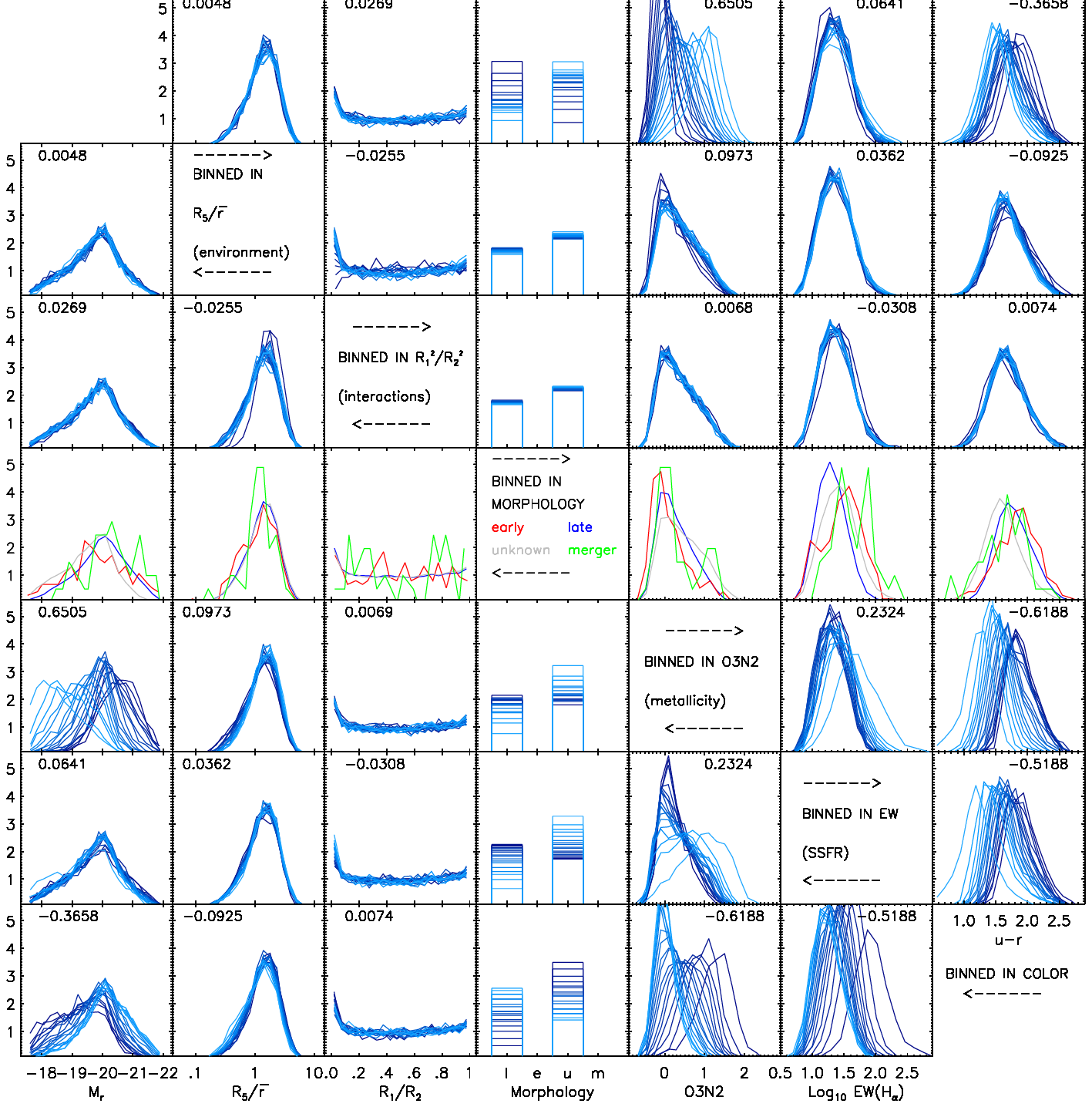}\\[-7mm]
\caption
{
From left to right, different columns show the distribution of mass ($M_{r}$), environment (\envir), interactions (\ratio), morphology (based on the Galaxy Zoo visual classification), metallicity (O3N2), and specific star formation rate, traced by EW(\Ha) and $(u-r)$ colour, for our subsample of \starforming\ galaxies.
Different lines correspond to adaptive bins in another of these properties (one per row) containing 2000 galaxies each.
Darker colours correspond to lower value bins, i.e. imply higher mass, denser environments, low values of \ratio, higher metallicity, lower EW, and bluer colour, respectively. 
}
\label{fig_metaplot}
\end{figure*} 
\section{Results}
\label{sec_results}

In order to study the physical mechanisms that regulate star formation, discriminating between `nature' and `nurture' processes, we present in this section a thorough characterization of the distribution and cross-correlations between stellar mass, environment, close companions, morphology, gas-phase metallicity, and specific star formation rate, using the observational proxies discussed in Section~\ref{sec_characterization}.

\subsection{Star-forming galaxies}

Our main results for the subsample of \starforming\ galaxies are summarized in Figure~\ref{fig_metaplot}, where each column shows histograms of $M_r$, \envir, \ratio, galaxy Zoo morphological classification, O3N2, equivalent width of the \Ha\ line, and $(u-r)$ colour.
In each row, different lines represent adaptive bins in the corresponding physical property, containing 2000 galaxies each.
Numbers within each panel indicate the Spearman's rank correlation coefficient between both variables.

\begin{figure*}
\centering
\includegraphics[width=.45\textwidth]{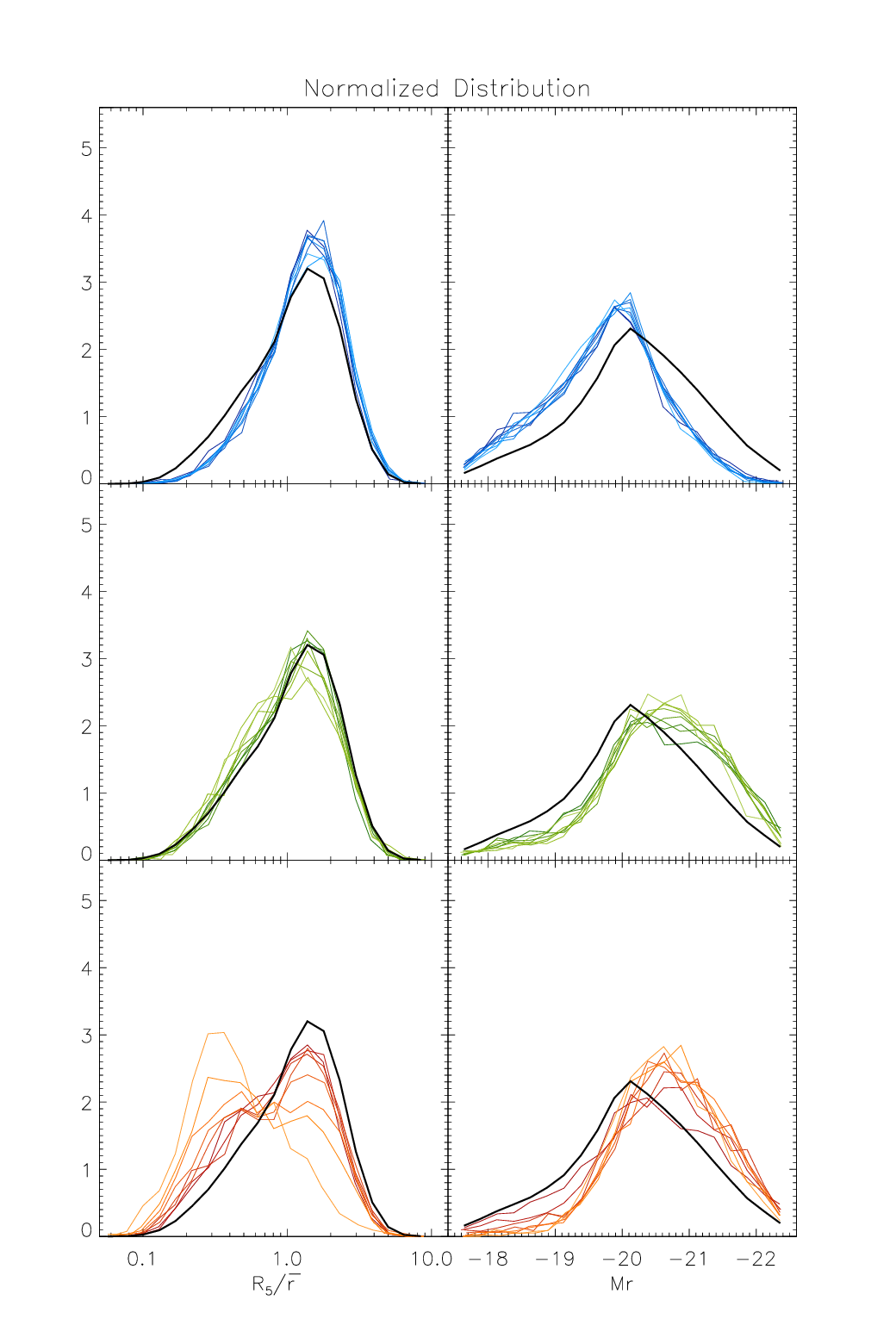}
\includegraphics[width=.45\textwidth]{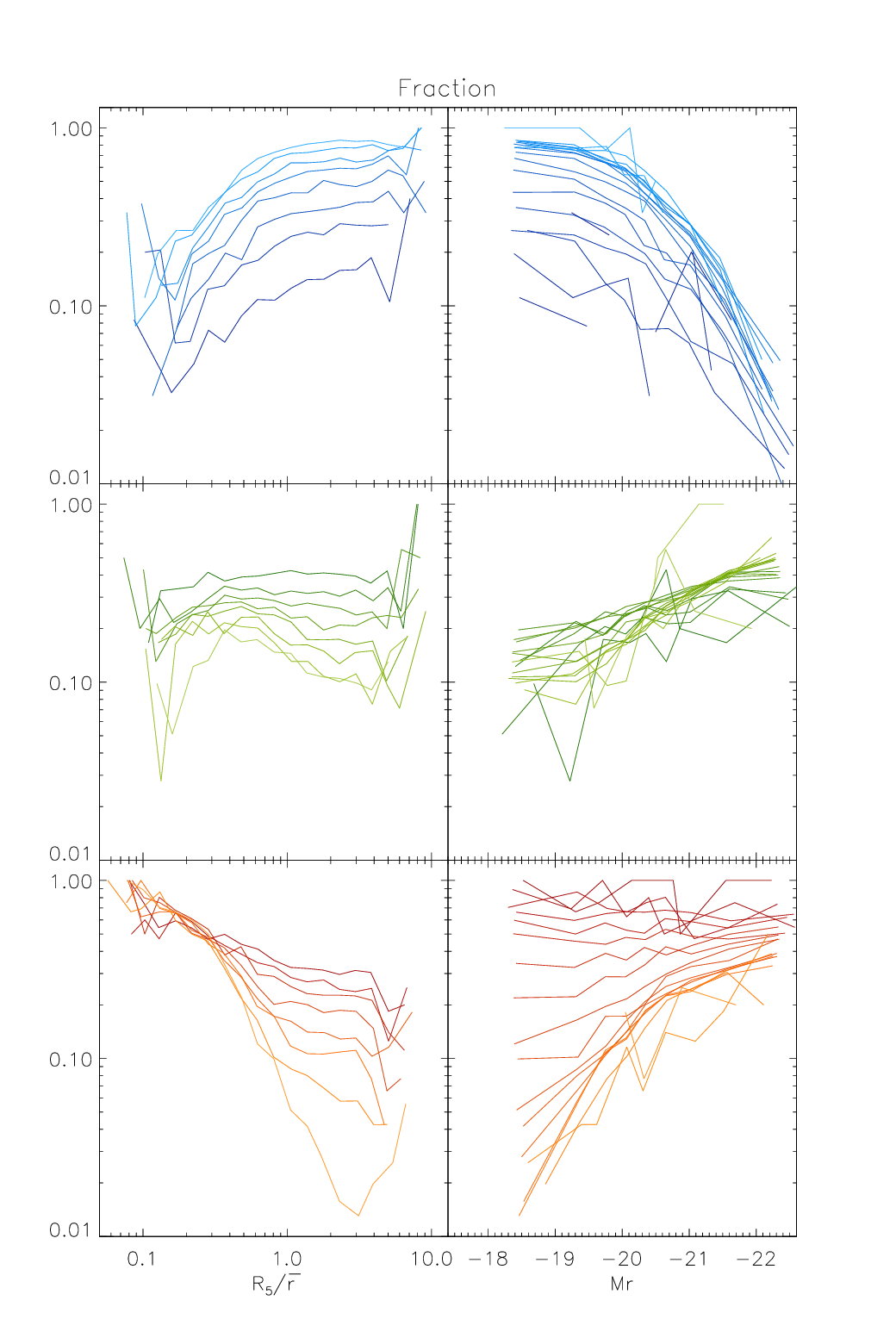}\\[-5mm]
\caption
{
Distribution (left panels) of \envir\ (environment, left) and $M_{r}$ (mass, right) for the \starforming\ (blue), the \intermediate (green) and 
the \passive (red) subsamples. The fraction of the three subsamples (right panels) is also represented against \envir\ and $M_{r}$. 
Line shading correspond to bins in the other variable and are fixed to contain 10000 galaxies. Darker colours correspond to lower value bins, i.e., more massive objects and denser environments. Black solid line in left panels correspond to the distribution of the whole sample.
}
\label{fig_environment}
\end{figure*}

When galaxies are binned by mass (top row), there are no apparent trends in the distributions of \envir\ and \ratio.
We find \citep[as others before, e.g.][]{Gomez+03} that \starforming\ galaxies tend to live in the field, and very few such objects are found in the densest environments.
Furthermore, we show here that this statement is not only valid in a qualitative sense; the distribution of \envir\ is independent on galaxy mass as long as the galaxy is classified as star-forming according to our criteria (see figure~\ref{fig_environment} for a comparison to the \intermediate\ and \passive\ populations).

In contrast, the well-known relation between metallicity and luminosity \citep{Lequeux+79} is clearly visible in the distribution of O3N2, with more luminous galaxies (darkest lines) displaying systematically lower values of O3N2 (higher metallicities).
There is a very weak trend in the distribution of equivalent widths (top row, sixth panel), in the sense of a more extended tail (i.e. a larger fraction of galaxies with higher SSFR, such as blue compact dwarfs and HII galaxies) in low-mass systems.
The trend is clearer when considering $(u-r)$ colour (rightmost panel), although the values of the rank correlation coefficients indicate that there is a weaker correlation between luminosity and colour or equivalent width than between luminosity and metallicity.

When galaxies are classified in terms of the environmental density (second row), we do not find any difference in the luminosity function of \starforming\ galaxies nor the distribution of their \Ha\ equivalent widths.
Nevertheless, we can observe that galaxies in the densest environments tend to be slightly more metal-rich (darkest lines in the second row, fifth column), in agreement with recent results 
\citep[e.g.][]{Hughes+13}, and display marginally redder colours.

Concerning the role of close companions, the distribution of \ratio\ (plotted on the panels in the third column) is roughly uniform, as expected for a random Poisson distribution in two dimensions.
However, there is a clear excess at $\ratio<0.1$ that we interpret as an indicator of interactions with the nearest neighbour.
The only difference between these galaxies (represented by the darkest lines in the third row) and the rest of the star-forming population is the lack of such close pairs in dense environments.
Since galaxies move at larger relative speeds in clusters than in the field, the typical type of two-body interaction is an impulsive encounter (`harassment') rather than the formation of a close pair and subsequent merger \citep[see e.g.][]{Boselli&Gavazzi06}.
There seems to be a very mild enhancement of the SSFR for the darkest bin ($\ratio<0.077$), observed in both \Ha\ and $(u-r)$ colour, but its statistical significance is rather weak.

The fifth row is binned in gas-phase metallicity, traced by the O3N2 line ratio.
One can see, as in the reciprocal panels in first and second rows, both the luminosity-metallicity relation and (to a much lesser extent) the slight deviation of the most metallic systems towards denser environments.
A clear correlation exists between O3N2 and SSFR, with lower-metallicity objects showing bluer $(u-r)$ colour and higher EW(\Ha).
In particular, about fifty per cent of the objects in the lowest metallicity bin \citep[with $\rm O3N2>1.25$, i.e. less than half solar according to][]{Perez-Montero&Contini09} feature values of the equivalent width higher than 50~\AA.

In the sixth row, where galaxies are binned according to their equivalent width, no evident correlation is found with galaxy density or the presence of close companions.
There is a marginal trend with luminosity, with the highest equivalent widths (larger than 65~\AA) appearing in systematically low-mass systems, as well as a significant correlation with O3N2 (albeit with a smaller correlation coefficient than the luminosity-metallicity relation).
Similar results are obtained using $(u-r)$ colour (seventh row) as an alternative proxy for the SSFR.
No correlation is found with our nurture indicators (\envir, \ratio), but there is a clear trend with galaxy luminosity.
The correlation with metallicity is even stronger than that observed for EW(\Ha).

Concerning morphological types, the fourth column of Figure~\ref{fig_metaplot} presents the fraction of galaxies classified as \emph{late-type}, \emph{early-type}, \emph{uncertain} (excluding mergers) or \emph{mergers}, binned in terms of each of the other proxies.
Conversely, the fourth row shows the distribution of each galactic property, segregated by our four morphological types (one can no longer build adaptive bins of 2000 galaxies).

Our \starforming\ subsample consists mainly on objects with late ($43.15$\%) and uncertain ($56.28$\%) morphologies, while early-type ($0.45$\%) and mergers ($0.12$\%) are barely present.
None of our `nurture' proxies (\envir\ and \ratio) seems to correlate with morphology, but we must recall that this refers to the \starforming\ subsample only.
The trends observed for the late-type and uncertain objects show that the former are, on average, more massive, more metallic, redder, and display lower values of EW(\Ha).
Early-type \starforming\ galaxies ($160$ objects) tend to display slightly higher metallicities, redder colours, and higher equivalent widths.
There are only 43 merging systems in the \starforming\ subsample, and therefore it is difficult to conclude whether there is a statistically significant increase in their average equivalent width.

\subsection{Separating `ageing' and `quenching'}

As shown in Section~\ref{sec_bias}, the signal-to-noise threshold imposed for our \starforming\ subsample is roughly equivalent to a selection criterion EW(\Ha)$>9$~\AA.
From Figure~\ref{fig_metaplot}, we have just concluded that there is only a slight dependence of the SSFR on galaxy luminosity, and almost no dependence on environmental density, when such a threshold is imposed.

We will now address the relation between galaxy mass, environment, and star formation, including also the \intermediate\ and \passive\ subsamples, thus covering all possible values of the \Ha\ equivalent width (with our adopted sign convention, negative values indicate a stellar absorption feature).
Once again, we first arrange all galaxies in bins (either in $M_r$ or \envir) containing 10000 objects.
Galaxies within each bin are then classified as \starforming, \intermediate, or \passive, as described in Section~\ref{sec_characterization}, and the respective fractions, as well as the conditional probability distribution of the physical property not used for the binning (\envir\ or $M_r$), are plotted in Figure~\ref{fig_environment}.
\emph{Star-forming} galaxies are shown on the top panels, whereas the central and bottom panels display \intermediate\ and \passive\ systems, respectively.

Let us start by focusing on the distribution of \envir, plotted on the leftmost column of the figure.
While \starforming\ and \intermediate\ galaxies tend to live in the field, \passive\ galaxies can be found in all kinds of environments.
In contrast with the other two populations, they also display a remarkably different behaviour at high and low luminosities.
At the low-mass end (light-shaded lines) they are mostly located in dense environments, but the distribution of \envir\ quickly becomes bimodal as we consider higher luminosity bins.
High-mass \passive\ galaxies (showing \Ha\ in absorption with $S/N>2$) can be found both in clusters and in the field, confirming the idea that there are indeed two mechanisms that may reduce (not necessarily `quench' on a short time scale) the star formation activity of a given galaxy to a negligible level.
One of these mechanisms would be related to galaxy mass (and, most probably, its `nature'), whereas the other, clearly related to the environmental density (i.e. `nurture'), would act upon galaxies of all masses \citep[see e.g.][]{Peng+10}.

As it has been previously reported in the literature, the fraction of star-forming galaxies (selected by either colour or \Ha) at fixed stellar mass is a strong function of local density, and there is also a clear dependence on galaxy luminosity at fixed \envir.
The fraction of galaxies classified as \starforming\ according to our criteria (top panels on the right column of Figure~\ref{fig_environment}) is indeed a monotonically decreasing function of both luminosity and galactic overdensity, although, contrary to some previous studies \citep[cf.][]{Tanaka+04}, we do not find evidence for a sharp transition at any particular value of either quantity.

Moreover, we would like to argue that the distinction between \passive\ and \intermediate\ galaxies is also a critical factor in order to discriminate between `ageing' and `quenching' mechanisms.
In low-density environments, most of the galaxies that are not \starforming\ fall in the \intermediate\ category, i.e. the fraction of truly \passive\ systems (\Ha\ unambiguously detected in absorption) is systematically lower than that of \intermediate\ objects for all luminosities.
As we will discuss later in more detail, the average SSFR of the field population is a decreasing function of stellar mass, but it seldom becomes negligible, especially for late-type galaxies; it merely tends to drop more and more often below the selection threshold defining the \starforming\ class.

As noticed by several authors \citep[e.g.][]{Baldry+06, Peng+10}, the effects of mass and environment on the fraction of \starforming\ galaxies are fully separable, and the curves with different shading on both of the top panels of the right column are parallel to each other.
However, these objects may gradually move towards the \intermediate\ population due to `ageing', staying there for a long time, or quickly migrate through the \intermediate\ class towards the \passive\ group due to a sharp quenching of their star formation activity.

If only a mass-independent quenching mechanism were at work, the fraction of \passive\ galaxies would increase at the expense of both \starforming\ and \intermediate\ objects in equal proportion, and therefore the ratio between their fractions would stay constant.
In contrast, one can readily observe that the effect of the environment on the fraction of \intermediate\ galaxies is not independent on mass.
The curves on the middle panels of the right column in Figure~\ref{fig_environment} are not parallel to each other, and they bear relatively little relation to the fraction of \starforming\ galaxies on the top panels.

Our results show that, as the local density increases, a certain fraction of the \starforming\ population becomes \intermediate\ and another undetermined fraction becomes \passive.
At the same time, a fraction of the \intermediate\ galaxies turns \passive, but they are approximately balanced by a similar number of \starforming\ galaxies moving into the \intermediate\ realm.
Incidentally, the final fraction of \intermediate\ galaxies turns out to be rather insensitive to the environmental density (middle right panel).

\begin{figure}
\centering
\includegraphics[height=.7\textheight]{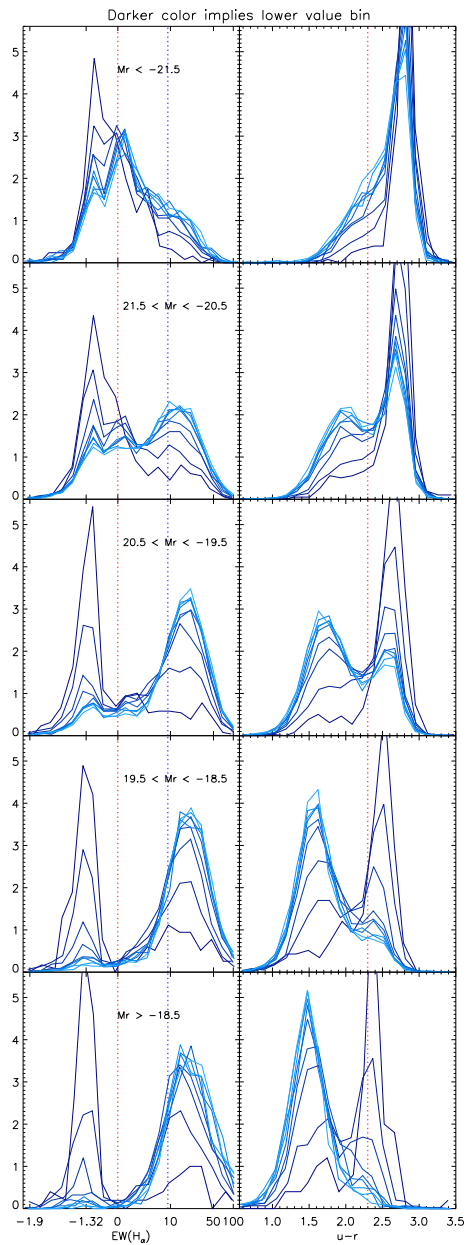}\\[-1mm]
\caption
{
Normalized distributions of EW(\Ha) (left) and $(u-r)$ (right) for different luminosity bins (brightest galaxies are plotted on the top panels).
Line shading indicates environmental density, with darker colours corresponding to the lowest values of  \envir\ (densest environments).
The binning in \envir\ has been computed from the whole target sample (10000 galaxies per bin), and it is the same in all panels.
}
\label{fig_EW}
\end{figure}


This process is further examined in Figure~\ref{fig_EW}, where we show the distribution of SSFR, traced by both colour and equivalent width, in different bins of mass \emph{and} environment.
We confirm the finding of \citet{Balogh+04_color} that the mean $(u-r)$ colour of the `blue' population (a mixture of \intermediate\ and \starforming\ galaxies) becomes progressively redder with increasing galactic overdensity.
A similar trend can be seen in EW(\Ha), whose distribution is plotted on the left panels.
As the density increases, the distribution becomes more asymmetric, with a tail extending into the \intermediate\ region, and its peak moves slightly towards lower values of the equivalent width.
While these effects are consistent with a mild `ageing' associated to denser environments, the dominant mechanism at play is a fast quenching of the star formation rate, responsible for quickly driving the objects towards the peak associated to the \passive\ population.

However, the role of \intermediate\ galaxies is not entirely clear in this context.
Are they transition objects, caught during the short interval of the transformation from \starforming\ into \passive\ after their SFR has suddenly dropped to zero, or will they live there for a long time, because their SFR has just decreased slightly (e.g. due to a shortage of external gas supply)?

In order to discern `ageing' from `quenching' (i.e. the time derivative of the star formation rate), it is interesting to compare two tracers of the SSFR that are sensitive to different time scales.
As recently shown by \citet{Schawinski+14}, early- and late-type galaxies occupy very different regions of the NUV$-u-r$ colour-colour diagram, with the former displaying systematically redder (NUV$-u$) colours, even for similar values of $(u-r)$.
Since UV emission probes star formation on shorter time scales (of the order of $10^7-10^8$~yr) than optical colours ($10^8-10^9$~yr), they concluded that early-type galaxies in the `green valley' have undergone a rapid end to star formation, whereas late-type galaxies with similar $(u-r)$ colours have experienced at most a slow decline in their star formation activity.

Rather than UV colours, we will use here the equivalent width of the \Ha\ line to probe the SSFR on scales of the order of $\sim 10^7$~yr.
Roughly speaking, the intensity of the \Ha\ line is proportional to the mass of O and B stars, whereas the continuum traces the total stellar mass.
Thus, EW(\Ha)$\sim M_7/M$, where $M_7$ denotes the stellar mass created during the last 10~Myr.
On the other hand, $(u-r)$ is sensitive to the presence of A stars, and it varies on scales of the order of 300~Myr.
To first order, one may consider that $(u-r)\sim M_{8.5}/M$, although this is only a coarse approximation (e.g. in the absence of dust extinction, no galaxy can be redder than a single stellar population with the current age of the universe).


\begin{figure*}
\vspace{-0.8cm}
\centering
\includegraphics[width=.95\textheight,angle=90]{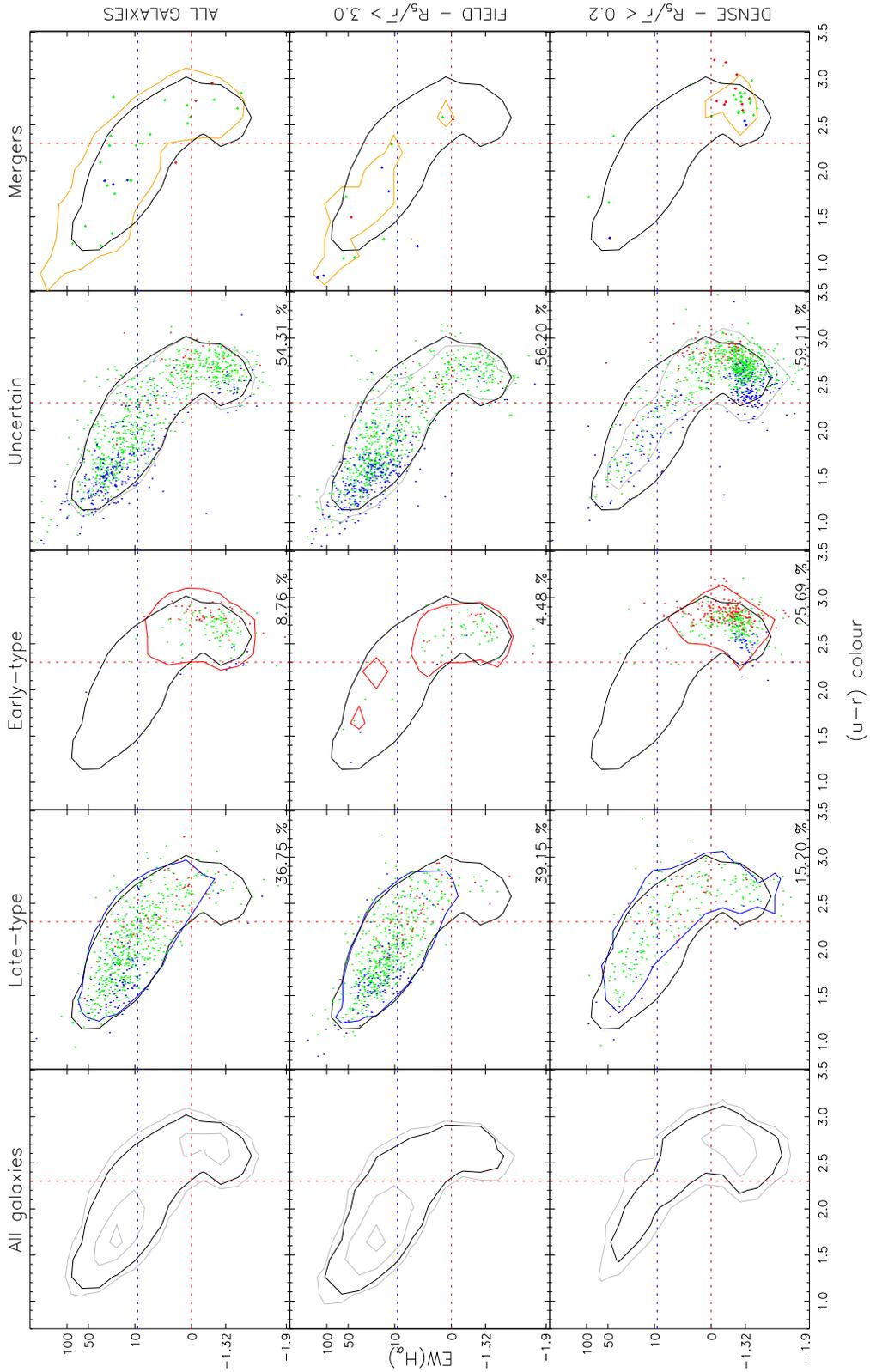}\\[-0.5cm]
\caption
{
Galactic colour-equivalent width diagram.
Top row corresponds to the whole galactic sample. Intermediate and bottom rows correspond to galaxies in low (field) and high density environments respectively.
From left to rigth (and in different columns), contours in black, blue, red, gray, and orange correspond to the area that encloses $90$\% of the \emph{total} galaxy sample, \emph{early-type}, \emph{late-type}, \emph{uncertain}, and \emph{merger} subsamples.
Points correspond to a random selection of target galaxies. We draw the same number of objects in the three rows coloured according to their \emph{mass} ($M_r$) in three different bins: red, objects with $M_r<-21.5$; green objects with $-21.5<M_r<-19.5$ ; and blue, objects with $M_r>-19.5$.
}
\label{fig_main_sequence}
\end{figure*}

If the star formation rate of a given galaxy evolved slowly ($\psi/\dot\psi>300$~Myr), there should be a one-to-one relation between its \Ha\ equivalent width and its $(u-r)$ colour, since both would be tracing exactly the same quantity; SSFR$=\psi/M$.
We will refer to this relation (the location of all galaxies with smooth star formation histories on the EW-colour diagram) as the `ageing' sequence.

If star formation were suddenly shut off, nebular \Ha\ emission would drop to negligible values on $\sim 10$-Myr scales, but the galaxy would still retain some memory of its original blue colour.
In fact, there would be a large amount of A stars and, therefore, recently quenched galaxies are expected to feature \emph{stronger} absorption lines (and bluer colours) than a genuinely old stellar population \citep{DresslerGunn83}.
The higher the SSFR before the quenching event, the stronger the \Ha\ absorption and the bluer $(u-r)$ after 10~Myr.
During the subsequent Gyr, A stars would gradually die; \Ha\ absorption will become weaker, and colours will become redder, until the galaxy eventually joins the red extreme of the `ageing' sequence.
Galaxies in this phase would occupy a completely different region of the EW-colour diagram, forming a well-defined line where EW(\Ha) is a monotonically \emph{increasing} function of $(u-r)$ that we will term the `quenched' sequence.

The $(u-r)$ colour -- EW(\Ha) diagram of our sample of SDSS galaxies is plotted in Figure~\ref{fig_main_sequence}, segregated by morphological type and environment, and coloured according to the object luminosity.
Although the ticks and labels in the $y$-axis indicate the actual value of EW(\Ha), the plotted quantity is in fact {log(EW/\AA+2)} in order to show both emission and absorption lines on a common logarithmic scale.

At the lowest overdensities, late-type galaxies seem to be well described by a single `ageing' sequence (i.e. $\psi/\dot\psi>300$~Myr).
The vast majority of them are forming stars at \emph{some} level, but they span a broad range of SSFRs, going from very active to very quiescent systems, and we do not detect any sharp transition whatsoever between the \starforming, \intermediate, and \passive\ populations.

Field early-type galaxies appear at the end of this sequence.
Some of them display weak \Ha\ emission, but almost none of them shows clear signatures (blue colours and strong \Ha\ absorption) of a recent quenching event.
Most of these systems are consistent with very low values of the SSFR (not necessarily of the instantaneous SFR) and a very old stellar population, in the sense of the mass-weighted average defined in equation~(\ref{eq_age}).

Galaxies with uncertain morphology are distributed along the whole sequence, but it is interesting to note that they reach the most extreme values at both ends (in fact, this category accounts for more than $80$\% of the galaxies with EW(\Ha)~$>65$~\AA).
From Figure~\ref{fig_main_sequence}, it becomes evident that galaxies showing the strongest quenching signatures also tend to display a disturbed appearance.
Interestingly, field galaxies unambiguously classified as ongoing mergers can only be found at the blue extreme.

In the densest environments, the `ageing' sequence defined by late-type galaxies moves toward redder colours and lower equivalent widths.
There is a general displacement of the sequence, consistent with the results plotted in Figure~\ref{fig_EW}, in the sense that galaxies of a given luminosity tend to be slightly older in dense environments.
However, the overall sequence stays roughly invariant (if anything, perhaps even narrower than in the field), with very few objects showing obvious post-quenching signatures.
Therefore, we conclude that late-type galaxies in dense environments have systematically lower SSFR and older stellar populations given their mass, but their star formation rate has not undergone any sudden change.

In contrast, many early-type galaxies in dense environments are arranged in a tight disposition with stronger \Ha\ absorption in bluer systems, consistent with the quenching scenario.
Furthermore, if we focus on galaxies in different luminosity bins, one can see them all over the `quenched' sequence, tracing the time elapsed since star formation stopped, but fainter objects, which typically started with higher SSFR, reach bluer colours and deeper absorption features than more massive systems.

The `ageing' and `quenched' sequences are also evident in galaxies with uncertain morphology, which constitute (as in the field) more than fifty per cent of the population.
They seem to reach bluer colours in the quenched sequence, but they do not particularly stand out among the ageing population.
A small number of mergers seem to be associated to objects with enhanced star formation, but most of them are located at the red extreme of the quenching sequence.

\begin{figure*}
\centering
\includegraphics[height=.45\textheight]{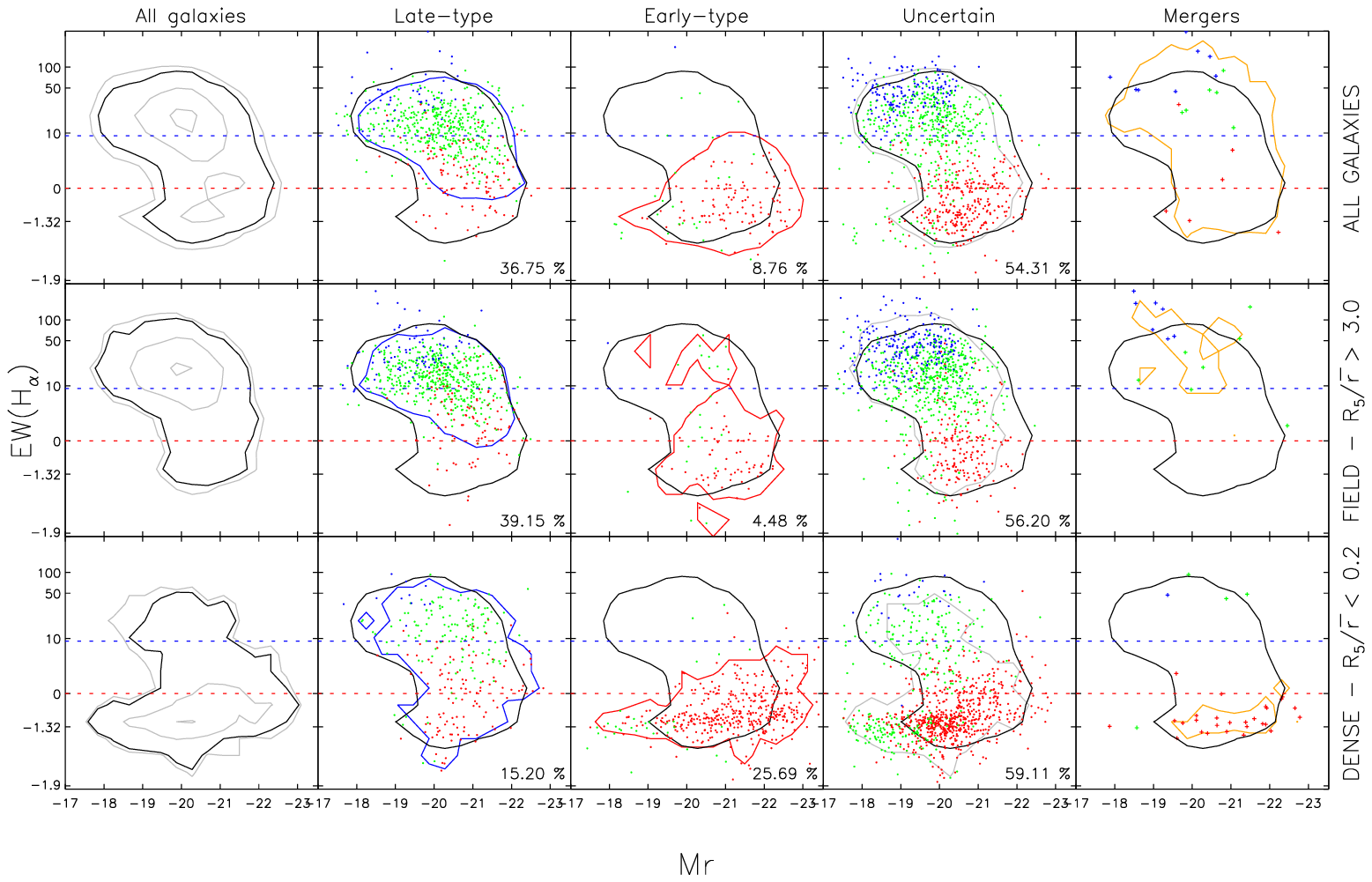}\\[-.1cm]
\includegraphics[height=.45\textheight]{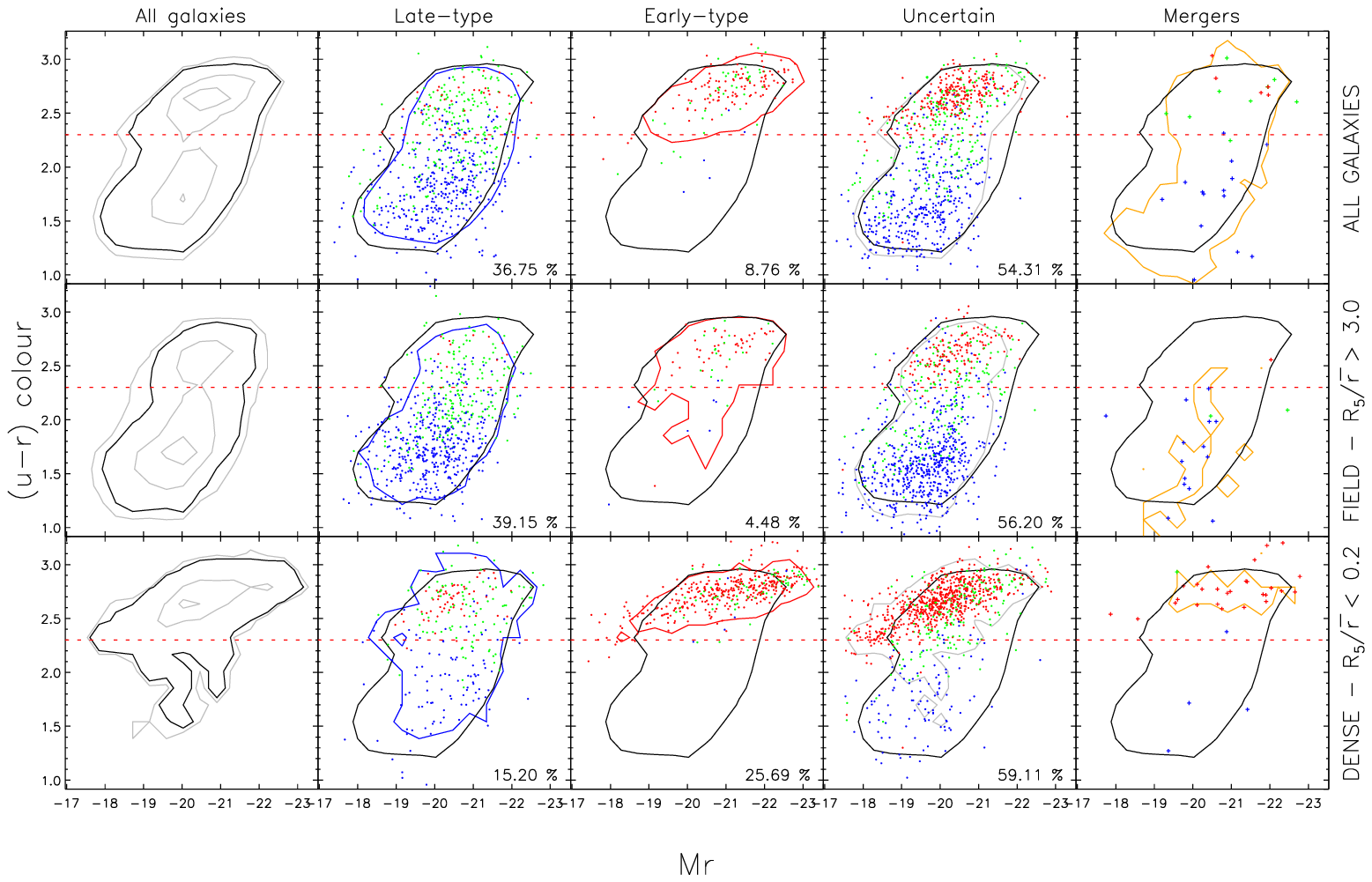}\\[0.1cm]
\caption
{
Galactic SSFR-luminosity relation, using \Ha\ equivalent width (top) and $(u-r)$ colour (bottom) as proxies for the SSFR. Figure schema is the same as in Figure~\ref{fig_main_sequence}.
The colours of the points correspond to:
Top -- three different bins in $(u-r)$ colour: red, objects with $(u-r)>2.5$; green objects with $1.5<(u-r)<-2.5$ ; and blue, objects with $(u-r)<1.5$.
Bottom -- three different bins in EW(\Ha): red, objects with EW(\Ha)$<0.0$; green objects with $0.0<$EW(\Ha)$<10.0$ ; and blue, objects with EW(\Ha)$>10.0$.
}
\label{fig_SSFR_L_relation}
\end{figure*}
\subsection{The SSFR-luminosity relation}

There is ample observational evidence that there exits a tight relation between the stellar mass and instantaneous/specific SFR of star-forming galaxies, often referred to as the `galactic main sequence' \citep[see e.g.][and references therein]{Noeske+07, Speagle+14}.
At low masses, the star formation rate is roughly proportional to the stellar mass, and therefore all star-forming galaxies display similar SSFRs, almost independent on luminosity.
The situation at high masses is still debated, but there is growing consensus that massive star-forming galaxies have consistently lower SSFR (albeit similar or higher SFR) than their smaller counterparts, especially in the local universe \citep[e.g.][]{Whitaker+14, Ilbert+14, Tasca+14}.

The SSFR of our galaxy sample, traced by either $(u-r)$ colour or EW(\Ha), is plotted in Figure~\ref{fig_SSFR_L_relation} as a function of galactic luminosity $M_r$, segregated by morphology and environment as in Figure~\ref{fig_main_sequence}.
In the EW(\Ha)$-M_r$ plane (top panels), galaxies are colour-coded by $(u-r)$, whereas in the colour-magnitude diagram (bottom panels), red, green, and blue correspond to different bins in equivalent width.

Since galaxies of early-type morphologies are known to display consistently redder colours than late-type spirals \citep[see the colour-magnitude diagrams on the bottom panels; cf.][]{Schawinski+14}, it is not surprising that they also feature lower equivalent widths, often (yet not always) showing the \Ha\ line in absorption (top panel).
Nevertheless, some early-type galaxies, especially in the field, display obvious signs of star formation, both in EW(\Ha) and in $(u-r)$.
As noted above, they seem to represent the low-SSFR end of the `ageing sequence' in Figure~\ref{fig_main_sequence}.
Figure~\ref{fig_SSFR_L_relation} suggests that, in the absence of environmental effects (i.e. in the filed) there is a single relation between SSFR and $M_r$.

We would strongly advocate for including \intermediate\ and \passive\ galaxies, of any morphological type, in the `galactic main sequence'.
As noted by \citet{Whitaker+14}, the shape of the sequence does not change by more than 0.1~dex when the whole galaxy population is considered.
Combining our Figures~\ref{fig_main_sequence} and~\ref{fig_SSFR_L_relation}, we do not find observational support for any quenching mechanism acting on field galaxies.
The most massive systems tend to have formed most of their stars a long time ago and have arranged them into an early morphological type, but there seems to be a continuous variation over the full range of SSFR.

In particular, our results strongly disagree with the two-population scenario, where some galaxies form stars with a given, mass-independent SSFR, until they are suddenly quenched \citep[e.g.][]{Peng+10}.
The apparent trend of a constant SSFR (as in e.g. Figure~\ref{fig_metaplot}) is entirely due to the selection of \starforming\ systems.
As shown in Figure~\ref{fig_comparison_SSFR}, our subsamples of \starforming, \intermediate, and \passive\ galaxies roughly correspond to fairly well defined cuts in colour or equivalent width.
In fact, they are statistically representative of the traditional `blue cloud', `green valley', and `red sequence' regions selected from the colour-magnitude diagram.
We argue that any such classification of the galaxy population introduces a serious bias on the average (S)SFR as a function of galaxy mass.

The situation in dense environments is completely different.
The stellar population of late-type galaxies becomes, on average, slightly older, and some of these objects do indeed belong to the \passive\ class.
At a given luminosity, the distribution of equivalent widths seems to broaden, reaching significantly lower values of EW(\Ha), whereas the $(u-r)$ colours tend to become somewhat redder, consistent with the `ageing' effect discussed in the context of Figures~\ref{fig_EW} and~\ref{fig_main_sequence}.
The fraction of elliptical galaxies increases dramatically with respect to the field, and many of them display signatures that are consistent with recent quenching.
Both statements are particularly evident in the case of low-mass galaxies, which typically display the highest SSFRs at low overdensities, and therefore the effect of the environment can be more clearly noticed as a sharp discontinuity between the galaxy properties of the `ageing' and `quenched' populations.

Galaxies with merging signatures display a clearly bimodal distribution.
In the field, they feature the highest SSFR, especially as far as the EW(\Ha) is concerned, whereas in dense environments they are systematically located at the red extreme, although it is not obvious whether they are more likely to belong to the `ageing' or the `quenched' sequence (cf. Figure~\ref{fig_main_sequence}).

\begin{figure*}
\centering
\includegraphics[height=.45\textheight]{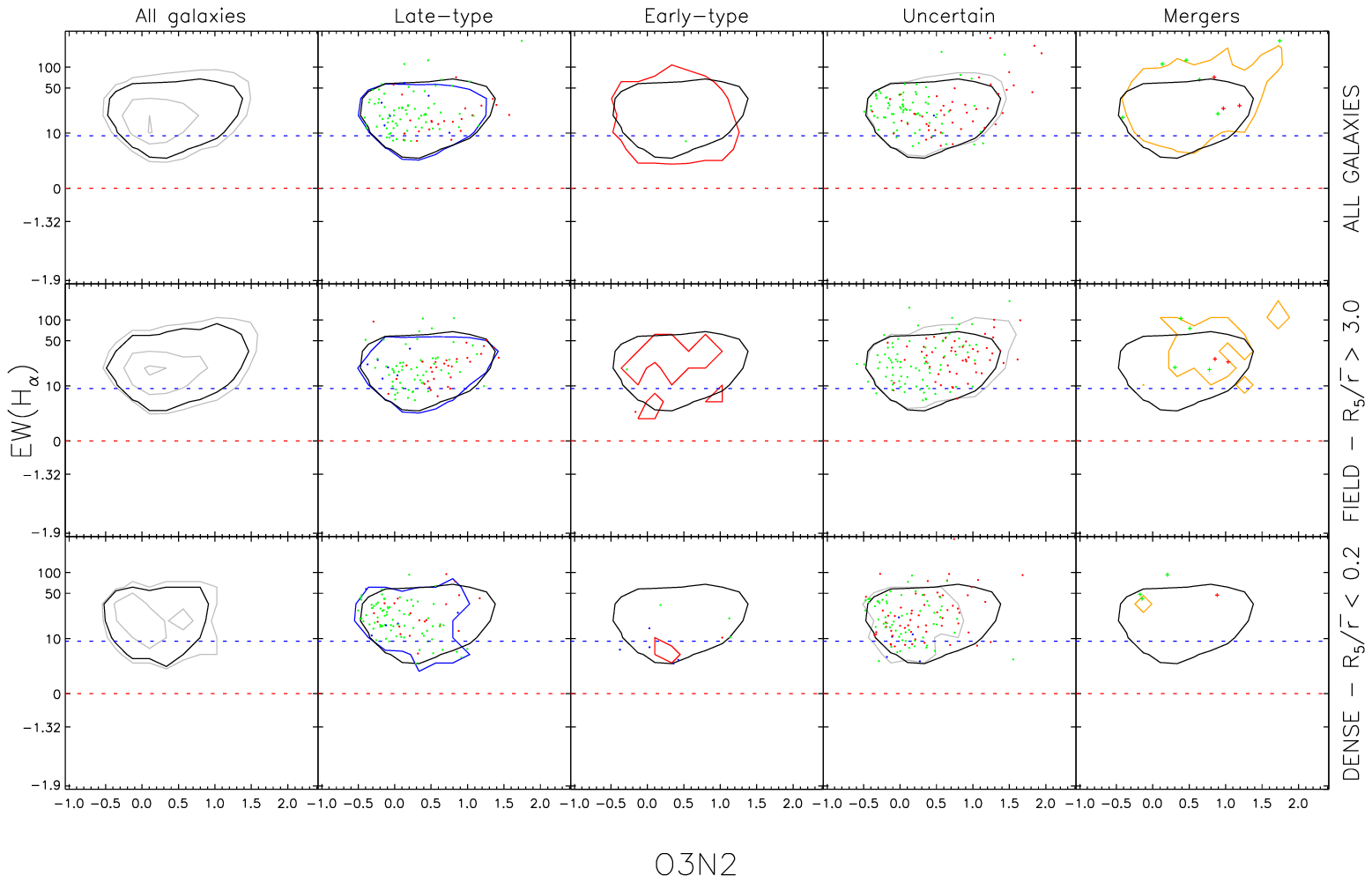}\\[-.1cm]
\includegraphics[height=.45\textheight]{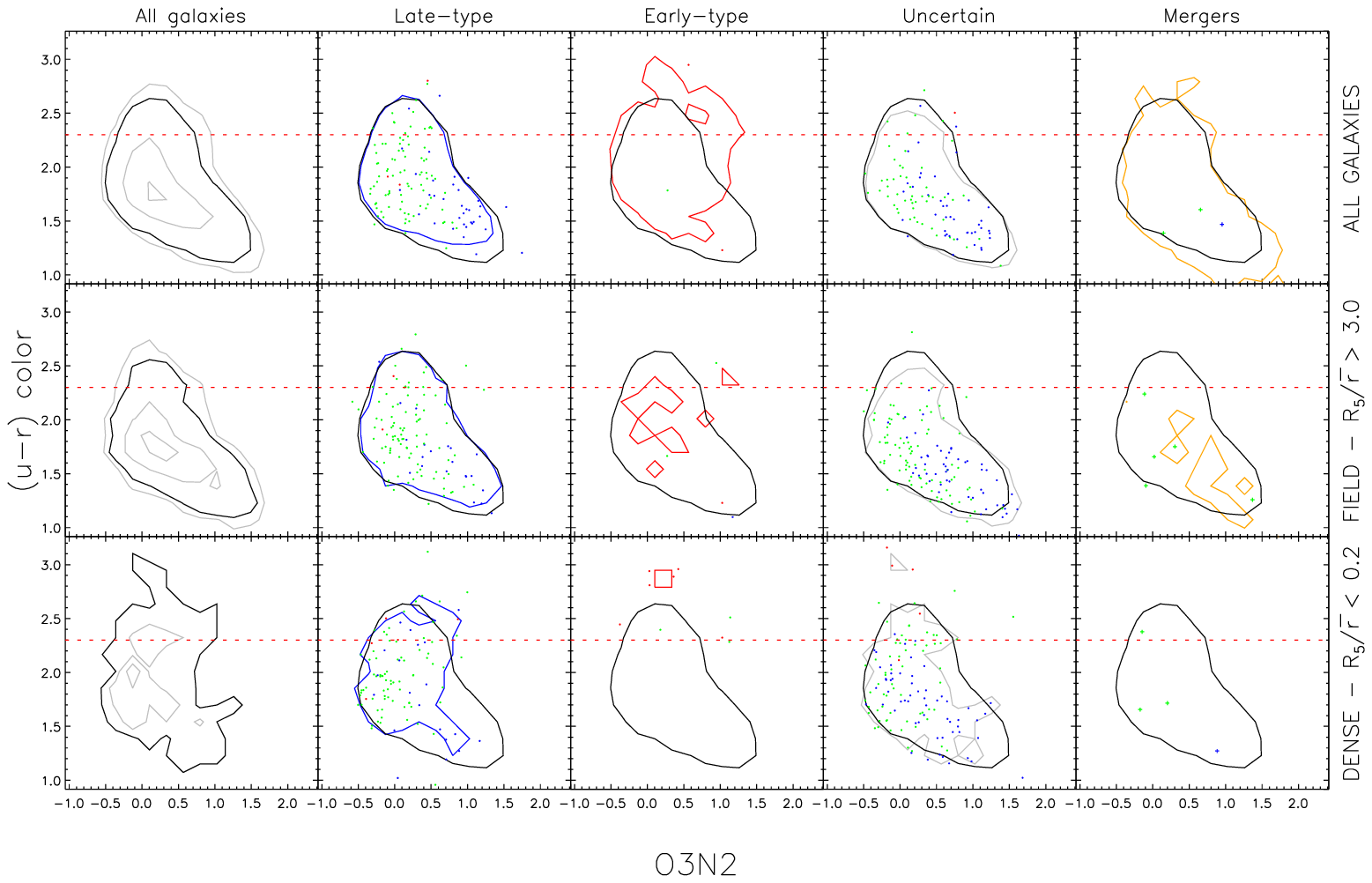}\\[0.1cm]
\caption
{
Galactic SSFR-metallicity relation, using \Ha\ equivalent width (top) and $(u-r)$ colour (bottom) as proxies for the SSFR, and O3N2 for the gas-phase oxygen abundance.
Figure schema is the same as in Figure~\ref{fig_main_sequence}.
The colours of the points correspond to:
both, top and bottom -- three different bins in \emph{mass} ($M_r$) : red, objects with $M_r<-21.5$; green objects with $-21.5<M_r<-19.5$ ; and blue, objects with $M_r>-19.5$.
}
\label{fig_SSFR_Z_relation}
\end{figure*}
\subsection{The SSFR-metallicity relation of star-forming galaxies}

Our results so far are consistent with a single `galactic ageing sequence' where most galaxies spend the largest part of their lives, including both passive and star-forming systems, with early- or late-type morphologies.
In addition to the correlations discussed in the present work, many other relations between the relative properties (colours, equivalent width, spectral shape, metallicity, gas fraction, mass-weighted mean age of the stellar population...) of a given galaxy have previously been reported \citep[see e.g.][]{Ascasibar&Sanchez-Almeida11,Ascasibar+14}, and therefore any of these parameters can be used to describe the current state of the galaxy in terms of its location along the sequence.

Here we will attempt to use the gas-phase oxygen abundance, traced by the O3N2 line ratio.
As shown in Figure~\ref{fig_metaplot}, the SSFR (measured in terms of colour or equivalent width) presents a higher correlation coefficient with O3N2 than it does with the absolute magnitude $M_r$.
On theoretical grounds, the gas-phase metallicity is expected to be closely related to the integrated star formation history of the galaxy, not only for the closed-box model \citep{Searle&Sargent72}, but for more elaborate scenarios including inflow and outflow of gas \citep[e.g.][]{Edmunds90}, even on resolved scales \citep{Ascasibar+14}.
The main disadvantage, of course, is that enough signal-to-noise ratio is necessary in order to estimate the O3N2 line ratio, and therefore we have restrained the discussion to the \starforming\ subsample.

The observed relation between the SSFR and O3N2 of \starforming\ galaxies, illustrated in Figure~\ref{fig_SSFR_Z_relation}, is indeed tighter than the relation between SSFR and luminosity discussed in the previous section.
As it was previously reported when discussing Figure~\ref{fig_metaplot}, the trend is clearer when using $(u-r)$ colour as a proxy for SSFR.
The SSFR-metallicity relation is fully compatible with `ageing' dominating the evolution of most \starforming\ galaxies, displaying almost exclusively late-type and uncertain morphologies.
Interestingly, we do see that the few mergers present in the sample do not only show, on average, higher SSFR, but also lower chemical abundance, than late-type or uncertain systems, hinting that they may be associated to the accretion of low-metallicity gas towards the main galaxy.
In all cases, galaxies with high SSFR (i.e. younger stellar populations) are metal-poor, and vice versa.
Additional work would be required in order to investigate whether the observed correlation between SSFR and metallicity is not only stronger, but indeed more fundamental than the correlation 
with stellar mass.

Regarding denser environments (recall here that we are missing \passive\ and \intermediate\ galaxies) we find that there is an evident lack of low-metallicity objects that could be explained either by an accelerated ageing (induced by e.g a reduction of the gas accretion rate with respect to the field) or by unidentified quenching mechanisms (e.g. stripping and/or rapid consumption) completely removing the gas from our systems and sending them to the quenched sequence.
In the latter case, it might be possible to find a quenching signature in the relation between SSFR and the mass-weighted stellar metallicity.

\section{Discussion}
\label{sec_discussion}

In this section, we would like to propose a scenario where a combination of both nature and nurture processes is responsible for regulating the SSFR of a given galaxy.
As noted in the introduction, we distinguish between smooth variations of the instantaneous/specific SFR (ageing), a sudden interruption of the star formation activity (quenching), and a transient episode of star formation that is sufficiently intense as to decrease the mass-weighted age of the underlying stellar population (rejuvenation burst).
Let us now briefly explore some possible physical mechanisms that may be responsible for these qualitatively different behaviours.

\subsection{Ageing}

In low-density environments, we argue that most galaxies should be expected to start their lives in a `chemically-primitive' (gas-rich, metal-poor, young stellar population, and high SSFR) state.
As they turn their gas into stars and enrich it with metals, they would gradually move towards a `chemically-evolved' (gas-poor, metal-rich, old stellar population, and low SSFR) state.
Let us stress once again that this `ageing' process is \emph{unavoidable} unless the star formation rate increases exponentially in time.
While possible, such sudden changes in the SSFR (see the subsection on `rejuvenation bursts' below) do not seem to dominate the overall evolution of typical field galaxies, although they may significantly perturb it for a limited amount of time and/or for a limited fraction of objects.

Considering the average properties of the galaxy population, unaffected by environmental factors, this work firmly establishes the existence of an `ageing sequence'; i.e. a fairly tight relation between the average EW(\Ha) and $(u-r)$ colour of unperturbed galaxies (extensive to O3N2, at least for the \starforming\ subsample).
This sequence is \emph{not} meant to represent chronological evolution, but a particular region of the parameter space where most galaxies are found \emph{at the present time} (and there is, in fact, no reason to expect that it should be invariant with cosmic epoch).
It merely reflects the relation that must exist between the physical properties of `young' and `old' galaxies if their star formation rate has varied slowly with time according to some smooth function $\psi(t)$, whose analytical description lies well beyond the scope of the present discussion.

Quite remarkably, dwarf galaxies (and, in particular, HII galaxies) tend to be located closer to the primitive extreme of the sequence, whereas more luminous systems tend to be, on average, more chemically evolved.
The physical origin of the relation between the total mass and the mass-weighted age of the stellar population (or the gas-phase metallicity) is an open question.
Ageing is related to the speed at which the gas is accreted and turned into stars, which may be affected by several `nature' and `nurture' processes.
The observed dependence on the environmental conditions may thus help us constrain the dominant physical mechanisms at work.

Galaxies in the field are consistent with a single trend, without any obvious discontinuity associated to the stellar mass or SSFR.
For `ageing' galaxies at fixed luminosity, there is a mild shift towards more chemically-evolved systems (higher metallicities and lower SSFR) in dense environments.
Since both the stellar mass and the galaxy overdensity are related to the mass of the host dark matter halo, this shift might be associated with `nature' processes driven by the relative efficiencies of gas infall, cooling, and consumption as a function of halo mass.
In addition, `nurture' processes, such as the strangulation of cold gas supply, may play a role in the accelerated `ageing' in dense environments.

As a final remark, let us note that galaxies with a smooth star formation history may be subject to morphological evolution, triggered by both internal and external mechanisms, and they may also be temporarily offset from the `ageing sequence' due to discrete quenching or rejuvenation events.
As long as they remain in the field, they will continue accreting gas and forming stars.
After some time, they will return to the `ageing sequence' and recover their original morphology, which may be most likely late-type or uncertain.

The time scales of such recovery will depend on the mass of the object, the availability of gas, and intensity of the event under consideration.
We argue that field elliptical galaxies, located at the end of the `ageing sequence', are representative of the last stages of secular galactic evolution.
We do not observe any convincing evidence that they have recently interrupted their star formation activity due to a merger event.
Rather, we propose that these galaxies, being more massive, already had a lower SSFR previous to such event, and therefore they kept their post-merger elliptical morphology up to the present day.
In this scenario, early-type morphologies would be a consequence, rather than the cause, of a reduced SSFR.

\subsection{Quenching}

One of the main results of the present work is that we only find unambiguous evidence for quenching in dense environments.
In these cases, star formation does not evolve gradually until the gas reservoir is exhausted.
On a very short time scale, these galaxies are removed from the `ageing sequence' by shutting down star formation to a level that \Ha\ is clearly detected in absorption.
Then, they evolve over the lifetime of A stars along a `quenched sequence' in the colour-equivalent width diagram, until both branches merge at the `red-and-dead' extreme.

The critical role played by the galactic overdensity hints that quenching is directly related to the mass of the dark matter halo.
It seems reasonable to expect that, for star formation to stop, it is necessary to prevent further accretion of cold gas.
This would be the reason why most galaxies in the field (in particular, low-mass galaxies) are almost invariably found among the `ageing' population.
Even if a discrete event might momentarily quench their star formation activity, it would be quickly resumed as a consequence of efficient gas accretion and cooling.

Nevertheless, the very existence of a `quenched sequence' implies that the physical process(es) responsible for environmental quenching must be very fast.
Preventing the accretion of new cold gas is a necessary but not sufficient condition; the galaxies' reservoir must be almost entirely consumed on a time scale much shorter than the lifetime of A stars.
Other `nurture' processes, such as ram-pressure compression and stripping, or tidal interactions and mergers with other galaxies, could contribute to quench star formation in dense environments, probably leaving a morphological signature.
From the top panel of Figure~\ref{fig_SSFR_L_relation}, we do indeed find that most of the low-mass galaxies featuring \Ha\ absorption lines have early-type or uncertain morphologies.
Some of them are classified as spirals, though, and it is difficult to assess whether the events (most likely nurture-related) responsible for the morphological transformation have also played a dominant role in the quenching of star formation, or, by the contrary, the observed morphology is a consequence of quenching, analogously to the scenario we proposed for field elliptical galaxies.

\subsection{Rejuvenation bursts}

Both `ageing' and `quenching' drive our systems towards a more `chemically evolved' state.
In particular cases (arguably much more common in the early universe), it is possible that some galaxies evolve towards more `chemically primitive' states for a limited period of time.
For instance, accretion of pristine gas well above the star formation rate would reduce the stellar-to-gas fraction and dilute the chemical composition.
It is also likely that such event drives a burst of star formation that would increase the metal abundance and stellar fraction, as well as the SSFR, yielding a higher equivalent width for the duration of the burst as well as higher luminosities and bluer colours for a longer time.
In extreme cases (regardless of gas accretion), a star-formation burst would decrease the mass-weighted average age of the stellar population.
This is our definition of `rejuvenation'.

Assuming that these processes leave an imprint on galaxy morphology, one can assess their statistical relevance on galaxy evolution by studying the correlations between colour, equivalent width, luminosity, and metallicity for galaxies displaying disturbed morphologies.
Our results clearly show that visually-identified mergers display systematically higher EW(\Ha) and lower $(u-r)$ for a given $M_r$ than galaxies with late-type morphology (or even those with uncertain morphology) for low-mass galaxies in low-density environments, i.e. those where a significant gas reservoir is expected.
We interpret this as evidence for merger-induced rejuvenation.

However, most galaxies with high SSFR have uncertain morphologies, which could well be associated to a recent merger, but they could also be due to intrinsic `nature' processes (e.g. stochastic star formation), especially in low-mass galaxies.
We can only ascertain that galaxies with irregular morphology have, on average, slightly higher SSFR than late-types, although the effect seems to be milder than the enhancement observed in ongoing mergers (which represent a very small fraction of the galactic population).
We argue that `rejuvenation' episodes contribute to the scatter observed in both, the `ageing sequence' and the relations between SSFR, metallicity, and luminosity. Nevertheless, these processes are not the dominant mechanism regulating neither the star formation nor the chemical evolution in galaxies.

It is possible, though, that they are important in particular types of objects, such as e.g. HII galaxies.
The presence of an underlying host with an older stellar population \citep[see e.g.][]{Papaderos+96, Amorin+09, Janowiecki+14} hints that these objects are good candidates for a `rejuvenation' episode.
However, it is not entirely clear whether the mass-weighted age of the stellar population, defined in equation~(\ref{eq_age}), has decreased as a consequence of a recent star formation burst or the SSFR has been a smooth function of time throughout most of their history.
Recent observations do indeed suggest that low-mass star-forming galaxies, including blue compact dwarfs, have formed 90 per cent of their stellar mass between 0.5 and 1.8 Gyr ago, consistent with the expectation from the SSFR-luminosity (or SSFS-metallicity) relation of `ageing' galaxies \citep{Rodriguez-Munoz+15}.

Since EW(\Ha) is sensitive to star formation $\sim 10$-Myr scales, we argue that close pairs or merger signatures should be clearly visible in the majority systems in the interaction-induced burst scenario.
Therefore, our results support the idea that HII galaxies are simply the `chemically primitive' extreme of the `ageing sequence', but we do not find evidence that their SSFR has increased drastically on very short time scales.

\section{Conclusions}
\label{sec_conclusions}

We have carried out an analysis of the mechanisms driving and regulating star formation in the local universe.
To do so we have selected a sample of galaxies from the SDSS DR7 spectroscopic catalogue, applying restrictions in apparent magnitude, redshift, and position in the sky, and then classified them as \starforming\ (35425 objects), \agn\ (8099 objects), \passive\ (19266 objects), or \intermediate\ (19826 objects) according to the observed intensity and signal-to-noise ratio of some of their optical spectral lines.

For the \starforming\ subsample, we have studied the distribution of absolute magnitudes in the $r$-band ($M_r$), the normalized distance to the fifth nearest neighbour (\envir), the ratio of squared projected distances of the first two neighbours (\ratio), the morphologies derived from the galaxy Zoo visual classification project, the O3N2 metallicity indicator, the equivalent width of the \Ha\ line, and the ($u-r$) colour.
Although we find evidence that mergers may temporarily increase the SSFR, we conclude that most (non-merging) \starforming\ systems evolve through internal processes, i.e. `nature', as we do not find, on average, any statistically-significant signature of the interaction-induced scenario.
We do find, however, that \starforming\ galaxies tend to be found in the field, and we show that the distribution of environmental density is in fact independent on other galaxy properties, including luminosity.
As far as we are aware, this is the first time such result has been reported in the literature.

Our analysis also shows that the restriction in signal-to-noise used to define our \starforming\ subsample, roughly equivalent to a fixed threshold in EW(\Ha) or $(u-r)$, creates an artificial bimodality when comparing \starforming\ and \passive\ galaxies.
Including the \intermediate\ population, we arrive at the following conclusions:

\begin{enumerate}

\item In low-density environments, most present-day galaxies are distributed along a relatively narrow `ageing sequence' in the EW(\Ha)-($u-r$) plane.
At the `chemically young' extreme, galaxies have high SSFR (blue colours and high equivalent widths), low metallicity (high O3N2), and late-type or uncertain morphology.
We argue that, at the `chemically old' extreme, galaxies tend to display elliptical morphologies because the SSFR is insufficient to rebuild a significant disk.
Brighter galaxies tend to be more chemically evolved, resulting in clear correlations between SSFR and luminosity or metallicity, but all of them are consistent with a smoothly-varying SSFR driven by secular (`nature') processes, with little evidence for recent `quenching' or `rejuvenation' episodes.

\item In dense environments, we detect a `quenched sequence' in the EW(\Ha)-($u-r$) plane, consistent with a very rapid truncation of the star formation activity.
Most of these objects display early-type or uncertain morphologies, but it is difficult to establish a causal connection between galaxy interactions and the drop in SSFR.

\end{enumerate}

We thus propose a scenario where `nature' is more important than `nurture' in regulating star formation in galaxies.
Starting their lives as `chemically primitive' objects, they gradually turn their gas into stars with a slowly-evolving SSFR and evolve towards `chemically old' systems, leading to the observed relation between EW(\Ha) and ($u-r$) colour.
Apart from this \emph{unavoidable} `ageing' process, galaxies are also susceptible to `quenching' (in dense environments) and/or `rejuvenation' episodes (associated to extreme bursts of star formation).
The ability to return to the `ageing sequence' after such events is linked to the external gas supply.
Galaxies will become \emph{red and dead} either when they totally exhaust their gas reservoir and stop forming stars (end of the `ageing sequence') or when they suffer a quenching event in a dense environment and they are no longer capable of accreting gas (`quenched sequence').
Mergers and galaxy-galaxy interactions may temporarily affect the instantaneous SSFR, but they merely seem to add statistical fluctuations to the main relation.

\section*{Acknowledgments}

This work has a long story, and its scientific scope has broadened considerably with respect to the original idea. It would have never been carried out without the enthusiastic collaboration of J. M. Domingo during the first stages of the project, and it would have not achieved its final form without the rigorous and constructive report on the first version by the anonymous referee.

Funding for the Sloan Digital Sky Survey (SDSS) and SDSS-II has been provided by the Alfred P. Sloan Foundation, the Participating Institutions, the National Science Foundation, the U.S. Department of Energy, the National Aeronautics and Space Administration, the Japanese Monbukagakusho, and the Max Planck Society, and the Higher Education Funding Council for England. The SDSS Web site is http://www.sdss.org/.

The SDSS is managed by the Astrophysical Research Consortium (ARC) for the Participating Institutions. The Participating Institutions are the American Museum of Natural History, Astrophysical Institute Potsdam, University of Basel, University of Cambridge, Case Western Reserve University, The University of Chicago, Drexel University, Fermilab, the Institute for Advanced Study, the Japan Participation Group, The Johns Hopkins University, the Joint Institute for Nuclear Astrophysics, the Kavli Institute for Particle Astrophysics and Cosmology, the Korean Scientist Group, the Chinese Academy of Sciences (LAMOST), Los Alamos National Laboratory, the Max-Planck-Institute for Astronomy (MPIA), the Max-Planck-Institute for Astrophysics (MPA), New Mexico State University, Ohio State University, University of Pittsburgh, University of Portsmouth, Princeton University, the United States Naval Observatory, and the University of Washington.

Financial support has been provided by projects AYA2010-21887-C04-03 (former \emph{Ministerio de Ciencia e Innovaci\'on}, Spain) and AYA2013-47742-C4-3-P (\emph{Ministerio de Econom\'{i}a y Competitividad}), as well as the exchange programme `Study of Emission-Line Galaxies with Integral-Field Spectroscopy' (SELGIFS, FP7-PEOPLE-2013-IRSES-612701), funded by the EU through the IRSES scheme.
YA is also supported by the \emph{Ram\'{o}n y Cajal} programme (RyC-2011-09461), currently managed by the \emph{Ministerio de Econom\'{i}a y Competitividad} (still cutting back on the Spanish scientific infrastructure).

 \bibliographystyle{mn2e}
 \bibliography{references}

\label{lastpage}
\end{document}